\documentclass[structabstract]{aa}

\usepackage{graphicx}
\usepackage{amsmath}
\usepackage{txfonts}
\usepackage{natbib}
\usepackage{hyperref}
\usepackage{xspace}
\usepackage{lineno}
\usepackage{rotating}
\usepackage{url}
\usepackage{gensymb}
\bibpunct{(}{)}{;}{a}{}{,} % to follow the A&A style
\hypersetup{
  colorlinks=true,
  citecolor=blue,
  linkcolor=blue
}
\newcommand{\uvot}{\textit{Swift}/UVOT}
\newcommand{\xrt}{\textit{Swift}/XRT}
\newcommand{\lat}{\textit{Fermi}/LAT}

\newcommand{\pks}{\object{PKS\,0048$-$097}}
\DeclareMathOperator\atanh{atanh}
\begin{document}

\title{Nine years of multi-frequency monitoring of the blazar PKS\,0048-097: spectral and temporal variability }

\titlerunning{Multi-frequency monitoring of PKS\,0048-097.}

\author{
  Alicja Wierzcholska\inst{1,2}.
}

\institute{
  \inst1Insitute of Nuclear Physics, Polish Academy of Sciences ul. Radzikowskiego 152, 31-342 Krak\'{o}w, Poland\\
  \inst2Landessternwarte, Universit\"at Heidelberg, K\"onigstuhl 12, D-69117 Heidelberg, Germany \\
  \email{alicja.wierzcholska@ifj.edu.pl}
}

\authorrunning{A. Wierzcholska}

\abstract
% context
{Blazars are highly variable, radio-loud active galactic nuclei  with jets  oriented at a small angle to the line of sight. The observed emission of these sources covers the whole electromagnetic spectrum from radio frequencies up to the high or even very high energy gamma-ray range. 
To understand the complex physics of these objects, multi-wavelength observations and studies on the variability and correlations between different wavelengths are therefore
essential.}
% aims
{The long-term multi-frequency observations of \pks\ are analysed
here to investigate its spectral and temporal features. The studies includes nine years of observations of the blazar, which is well studied in the optical and radio domain, but not in the other frequencies.}
% methods
{Multi-wavelength data collected with OVRO, KAIT, Catalina, \uvot\, \xrt\ and \lat\ were studied.}
% results
{The performed analysis revealed strong variability in all wavelengths
that is most clearly manifested in the X-ray range. The correlation studies do not exhibit any  relation between different wavelengths, except for the very strong positive correlation between the optical emission in V and R bands.}
% conclusions
{}


\maketitle

\section{Introduction}
Blazars constitute an extreme class of radio-loud active galactic nuclei (AGN), which are characterized by a relativistic jet that
is pointed at small angles to the observer \citep[e.g.][]{begelman84}, and polarized and highly variable non-thermal continuum emission extending through the entire electromagnetic spectrum, from radio to X-rays and, or even, up to high and very high energy $\gamma$-rays
\citep[e.g.][]{Gupta, Wagner2009, Vercellone, Giommi, Abramowski2013, Abramowski2014}. The  
rapid variability of blazars is  visible at different wavelengths on different timescales down to hours or even shorter \citep[e.g.][]{wagner,2155flare,saito}. 

This  class of objects includes the flat-spectrum radio-loud quasars (FSRQs) as well as the BL Lacertae type (BL Lac) objects. For
FSRQs the presence of prominent broad and narrow emission lines are characteristic, while the featureless continuum emission in the optical band is  attributed to the BL Lac type objects \citep[e.g.][]{urry}. 
The spectral energy distribution (SED), in $\nu F_{\nu}$ representation, usually has a double-peaked structure that is generated by two emission components. 
The first bump located between the optical and X-ray regimes is described by the synchrotron radiation from the relativistic electrons in the jet. The second bump can be explained either by leptonic or by the hadronic scenarios. 
In the leptonic models the high-energy part of the SED, located in the hard X-ray-to-gamma-ray regime, is produced by the inverse-Compton (IC) emission of the same electron 
population, involving either the jet synchrotron photons as a seed for the IC scattering 
\citep[synchrotron self-Compton model, SSC; e.g. ][]{konigl, marscher, Band85}, or various photon 
fields originating outside of the jet \citep[external-Compton models; e.g. ][]{dermer92, 
sikora}.
Alternatively, the second bump can be explained in the framework of hadronic scenarios that are mostly initiated by the relativistic protons accelerated with the electrons \citep[see e.g.][]{Mannheim92, Mucke03, boe07}.
The location of the low-energy peak in the SED subdivides blazars into three subclasses of high-, intermediate-, and low-energy peaked sources (HBLs, IBLs, and LBLs, respectively), depending on the position of their synchrotron peak frequencies \citep[see e.g.][]{padovani95,fossati98, Abdo2010}.

\pks\ (RA$_{2000}$: $00^h50^m41.317^s$, DEC$_{2000}$: $-09\degree 29' 05.21''$) is a BL Lac-type blazar \citep{Plotkin08}  located at redshift $z=0.635$ \citep{Landoni12}, which is well studied in the optical and radio regime \citep[e.g.][]{Ross70, Carswell73, Pica88, Wills92, Falomo96, Stickel93, Sefako01}. 
The source was reported in the first \citep{1fgl} and second \citep{2fgl} Fermi-LAT Catalogs and Fermi Bright Gamma-ray Source List \citep{0fgl}, known as 1FGL, 2FGL and 0FGL, respectively. No detection in the very high energy gamma-rays range has been reported, meaning that there is only an upper limit on flux provided by the H.E.S.S. Collaboration \citep{hess_up}. 
\cite{Abdo2010}  found the frequency of lower peak to be $\log \nu_{syn}=14.3$, which allows classifying the object as an IBL-type source. 

For the first time, we present here the results of nine years of multi-wavelength observations of \pks\ covering the radio, optical, UV, X-ray, and high-energy gamma-ray wavelengths.
The paper is organized as follows: Sect.~\ref{data} describes the multi-frequency observations of the source, Sect.~\ref{behaviour} shows the behaviour of \pks\ in both the quiescence and flaring states and the spectral properties in GeV range. Section~\ref{sec:correlations} discuss the multi-frequency correlations. The work is summarized in Sect.~\ref{summary}.

\section{Multi-wavelength data} \label{data}

\subsection{Gamma-ray monitoring with Fermi/LAT}
The LAT, on-board the \textit{Fermi} satellite, is a pair-conversion detector that is sensitive to photons in the  energy band from $\sim 20$\,MeV to a few hundred GeV \citep{atwood}. This primary mission instrument covers the full sky every three\,hours, and  the data are available publicly on the mission web page\footnote{\url{http://fermi.gsfc.nasa.gov/ssc/data/access/}}. 

\pks\ is included in 2FGL (as well as in previous catalogues: 0FGL and 1FGL) as  2FGL\,0050.6$-$0926 with an average flux between 100\,MeV and 100\,GeV of $(3.85 \pm 0.25)\cdot10^{-9}$\,ph\,cm$^{-2}$\,s$^{-1}$ and the spectral index of $\Gamma_\textrm{2FGL} = 2.14 \pm 0.04$. 
For comparison, the 1FGL reports an average flux in the same energy range of $(4.50 \pm 0.45) \cdot 10^{-9}$\,ph\,cm$^{-2}$\,s$^{-1}$ and a spectral index of $\Gamma_\textrm{1FGL} = 2.19 \pm 0.05$, while the reported flux level in 0FGL is $(7.2 \pm 1.0) \cdot 10^{-9}$\,ph\,cm$^{-2}$\,s$^{-1}$.

For this paper data collected between August 4,$^{}$ 2008 and July 1,$^{}$ 2014 were analysed using standard \textit{\textup{Fermi Science tools}} (version v9r33p0) with \verb|P7REP_SOURCE_V15_rev1| instrument response functions (IRFs). For the analysis the photons with a zenith angle $<105^\circ$ were selected in the energy range of 100\,MeV to 300\,GeV. Events were selected in a $15^\circ$ region of interest (ROI) centred on \pks\ .
The binned maximum-likelihood method \citep{Mattox96} was applied in the analysis.
The Galactic diffuse background was modelled using the \verb|gll_iem_v05| map
cube, and the extragalactic diffuse and residual instrument backgrounds were
modelled jointly using the \verb|isotropic_iem_v05| template. All the sources from the {\it Fermi}-LAT Second Source Catalog 
\citep[2FGL,][]{2fgl} inside the ROI of \pks\ were modelled. 

The long-term light curve binned in seven-day intervals is presented in Fig.~\ref{all_lc}.

\subsection{X-ray  observations with \textit{Swift}}
The \textit{Swift} mission \citep{Gehrels04} is a multi-wavelength space observatory launched into orbit on November 20$^{}$, 2004. The instrument is equipped with the following detectors: the Burst Alert Telescope \citep[BAT,][]{Barthelmy05}, the X-ray Telescope \citep[XRT,][]{Burrows05}, and the Ultraviolet/Optical Telescope \citep[UVOT,][]{Roming05}.

\pks\ was monitored with \xrt\  in nine pointed observations (all in PC mode), resulting in total exposure of 32.1\,ks. 
These data were analysed using version 6.15 of the HEASOFT package\footnote{\url{http://heasarc.gsfc.nasa.gov/docs/software/lheasoft}}. Data were recalibrated using the standard procedure \verb|xrtpipeline|. All the observations were checked for the pile-up effect, which was found to be negligible. 
Spectral analysis was performed for data in the energy range of 0.3-10\,keV with the latest version of the \verb|XSPEC| package (version 12.8.2).
All data were binned to have at least 30 counts per bin. Spectra are well fitted with a power-law function with a Galactic absorption 
 value of $N_{H} = 3.22 \cdot 10^{20}$\,cm$^{-2}$ \citep{Kalberla05} set as a frozen parameter. 
 The power-law fit parameters are collected in Table~\ref{table_xrt}.

The long-term light curve of the integrated flux in the energy range  of 2.0-10\,keV,  presented in Fig.~\ref{all_lc}, shows the significant variability of the source. There are no significant changes in the spectral index for the period of observations (see Table~\ref{table_xrt}). 

\subsection{UV observations with \textit{Swift}}
The UVOT instrument measures the UV and optical emission simultaneous to the X-ray telescope.
The observations are taken in the UV and optical  bands with 
central wavelengths of UVW2 (188 nm), UVM2 (217 nm), UVW1 (251 nm), U (345 nm), B
(439 nm), and V (544 nm). 

The instrumental magnitudes were calculated using \verb|uvotsource| taking into account all
photons from a circular region with radius 5''. The background was determined from a circular region with radius 10'' near the source region. 
The flux conversion factors used are provided by \cite{Poole08}. 
All UVOT data were corrected for the dust absorption using the reddening $E(B-V)$ = 0.0274\,mag  \citep{Schlafly} and the ratios of the extinction to reddening, $A_{\lambda} / E(B-V)$, for each filter \citep{Giommi06}.
All the measured magnitudes are collected in Table~\ref{table_uvot}, while the
\uvot\ light curve is presented in  Fig.~\ref{all_lc}.

\subsection{Optical monitoring with Catalina and with KAIT}
\pks\ is one of the frequently monitored targets by the two optical instruments Catalina and KAIT, which observe the blazar in V and R bands, respectively. Both light curves were used in this paper. 

The Catalina  Survey \cite[][]{Drake09} consists of the Catalina Sky Survey (CSS) and the Catalina Real-time Transient Survey (CRTS).
Here only CSS data in V band are used, which are publicly available on the instrument web page\footnote{\url{http://nesssi.cacr.caltech.edu/catalina/Blazars/Blazar.html}}.

The second optical monitoring we present was obtained with The Katzman Automatic Imaging Telescope Gamma-Ray Burst
KAIT \citep{Li00}. KAIT is the third robotic telescope in the Berkeley Automatic Imaging Telescope
(BAIT) program \citep{Richmond93,Treffers95}. The instrument monitors 163 AGNs with an average cadence of three days.
Data points in the light curves in R band \citep{Li03}  are produced through a pipeline described by \cite{Cohen14} and available on the program web page\footnote{\url{http://brando.astro.berkeley.edu/kait/agn/}}. 

The host galaxy of \pks\ remains unresolved \citep{Kotilainen98}, and because of this, 
 the data were not corrected for the influence of the host.
All optical magnitudes were corrected against the Galactic extinction based on the model by \cite{Schlegel98} with the most recent recalibration by \cite{Schlafly}, using the correction factors of  $A_V=0.088$ and  $A_R=0.069$, for V and R band, respectively.

\subsection{Radio observations with OVRO}
The radio observations of \pks\ were carried out at 15\,GHz with the Owens Valley Radio Observatory (OVRO), which is the 40\,m telescope dedicated to observe \emph{Fermi}-LAT targets \citep{Richards11}. Data used in this analysis were collected between January 6,$^{}$ 2008 to May 8,$^{}$ 2014 and were downloaded from the programme website\footnote{\url{http://www.astro.caltech.edu/ovroblazars}}. 

%--------------------------
\begin{figure*}[t]
\centering{\includegraphics[width=0.98\textwidth]{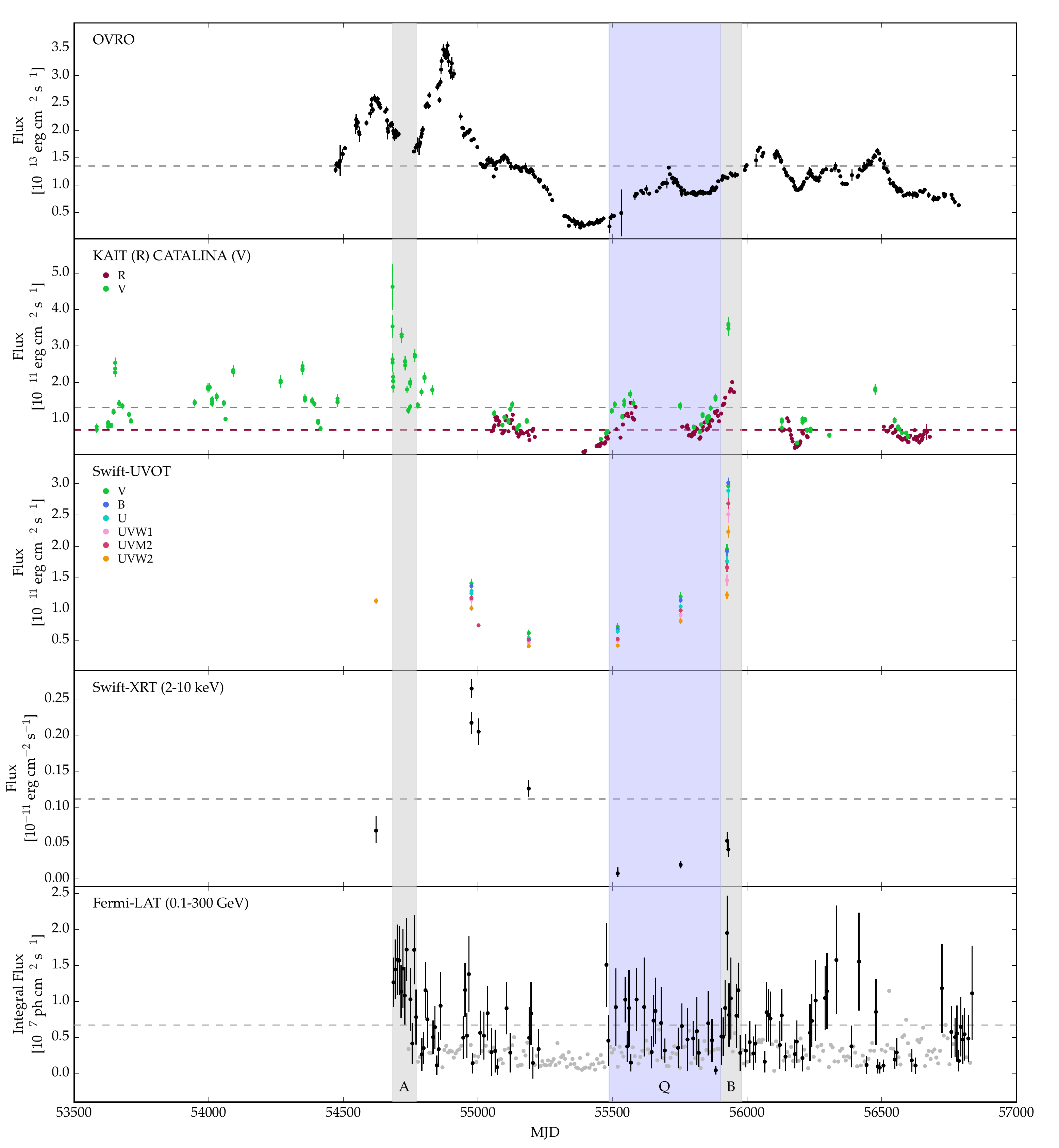}}
\caption[]{Multi-wavelength light curve of \pks. Panels from the top to the bottom show the radio observations from the OVRO telescope at 15\,GHz; optical (KAIT and Catalina) monitoring in R and V band, Swift/UVOT data in V, B, U, UVW1, UVM2, and UVW2 filters, X-ray observations with Swift/XRT, and Fermi-/LAT flux ($E>100$\,MeV). The horizontal dashed lines in all the panels represent the average flux for all observations presented here. In all the cases (excluding Fermi/LAT data) each point corresponds to one night of observations; for the case of Fermi/LAT monitoring data are binned in week-long intervals. 
In the Fermi/LAT light curve, flux upper limits are represented with grey points.
The time periods corresponding to A and B flares are marked in grey, while the quiescence period Q is plotted in blue.}
\label{all_lc}
\end{figure*}
%--------------------------

\section{Multi-wavelength behaviour and flaring activity of PKS\,0048$-$097} \label{behaviour}
Studies on multi-wavelength behaviour using simultaneous observations of the blazars, both in quiescence and flaring state, are crucial to understand the physics and nature of these objects.
\pks\ is a highly variable source, significant flux changes are observed in all wavelengths (see Fig.~\ref{all_lc}).

\subsection{Temporal and spectral variability of the blazar}
\begin{figure}[t]
\centering{\includegraphics[width=0.50\textwidth]{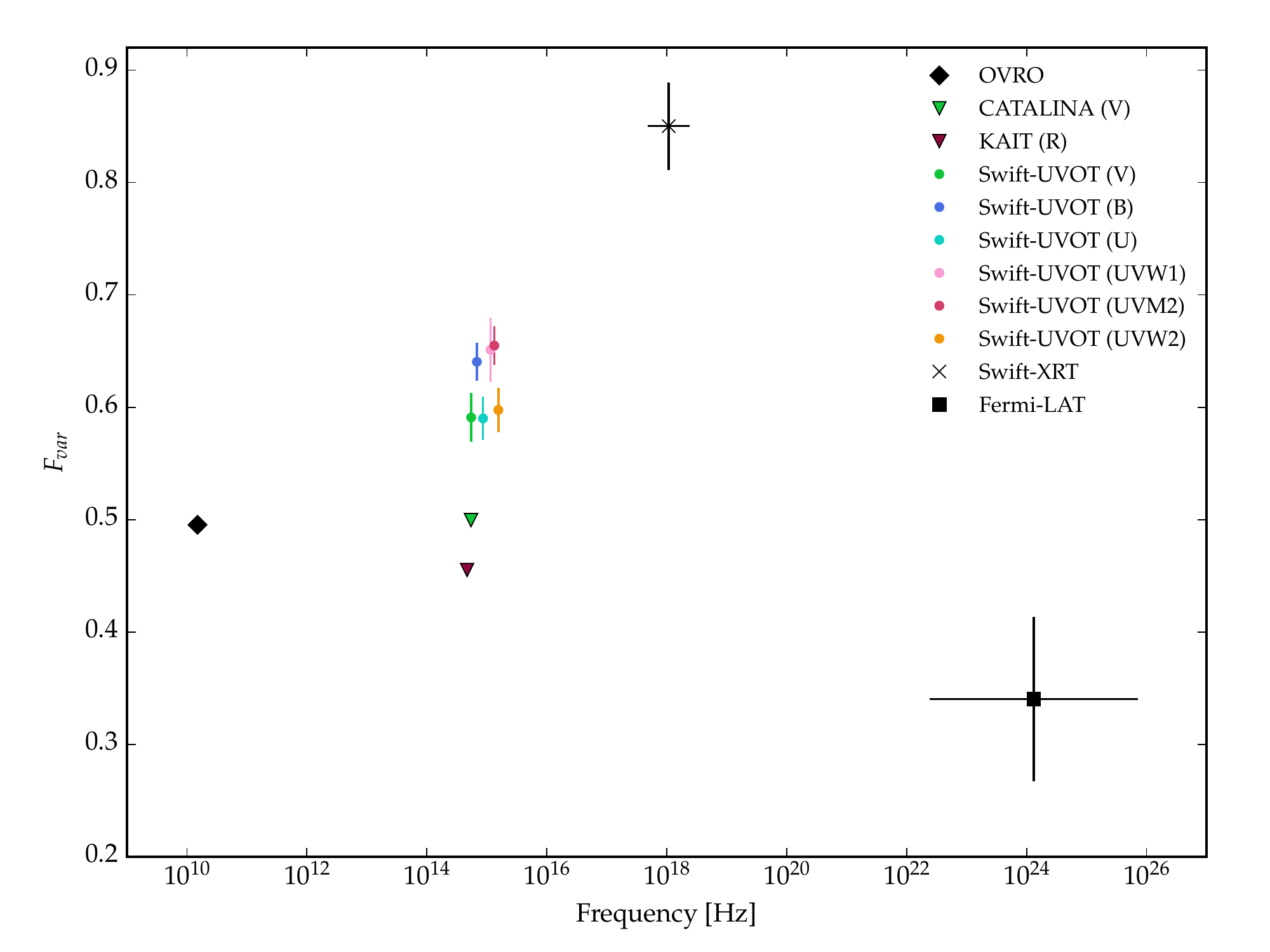}}
\caption[]{Fractional variability vs. frequency for each observation regime. The numerical values can be found in Table~\ref{table_fvar}. The colours for the data points are the same as in Fig.~\ref{all_lc}.}
\label{fvar}
\end{figure}

To quantify the temporal variability that is observed in different wavelengths, the fractional variability amplitude is calculated using formula provided  by \cite{Vaughan03},

\begin{equation}
 F_{var}= \sqrt{\frac{S^2-e^2}{F^2}},
\end{equation}
where $S^2$ is the variance, $e^2$ is the mean square error, and $F$ is the mean flux. 
The uncertainty of $F_{var}$ is calculated following the formula by \cite{Poutanen08},

\begin{equation}
  \delta  F_{var}= \sqrt{F_{var}^2+(\sigma^2)} -F_{var},
\end{equation}
with the error in the normalised excess variance $\sigma$ given as \citep{Vaughan03}
\begin{equation}
 \sigma = \sqrt{\left( \sqrt{\frac{2}{N} }\frac{e^2}{F^2} \right)^2   + \left( \sqrt{\frac{e^2}{N}}\frac{2 F_{var}}{F} \right)^2        },
\end{equation}
where $N$ is the number of data points in the light curve. 
The calculated values of $F_{var}$ for all the energy bands we
analysed are collected in Table~\ref{table_fvar}.
Figure~\ref{fvar} shows changes of the $F_{var}$ values in different energy bands.
During the whole monitoring period of the blazar, significant variability was revealed in the  $F_{var}$ values. The lowest $F_{var}$ value is for the GeV observation, the highest for X-ray range,  while for the other ranges the $F_{var}$ values are similar.  
We note that for \lat\ data $F_{var}$ was only calculated for
the flux points, and the upper limit points were omitted from the calculations.

It is important to mention that the values of $F_{var}$ are strongly dependent
on the size of the time bins in the light curves.  Smaller bins allow showing
stronger flux variations, which in the case of larger time bins can be smoothed out
and lead to lower $F_{var}$ values.  Obviously, the time binning is limited by
the characteristics of particular instruments.
The second factor that influences the values of $F_{var}$ are the flux
uncertainties. The uncertainties according to the definition are expected to be
constant (or at least very close to constant), which in practice is not always
true.  For example, such a case is visible in the optical monitoring with
Catalina (see Fig.~\ref{all_lc}), in which some measurements have much larger
uncertainties than the others.

Figure~\ref{all_lc} reveals the variable nature of the source. Nine years of this this long-term monitoring allow pointing out significant variability patterns that are observed in different wavelengths. 

\subsection*{Radio variability}
In the radio band the flux oscillates between $0.2\cdot10^{-13}$\,erg\,cm$^{-2}$\,s$^{-1}$ and $3.6\cdot10^{-13}$\,erg\,cm$^{-2}$\,s$^{-1}$. 
The highest mentioned value corresponds to about 2.4\,Jy,  the lowest to about 0.2\,Jy.
The highest flux level is observed during the period of MJD54762$-$MJD54998, while the lowest values are collected during the period of MJD55321$-$MJD55505.
As mentioned before, \pks\ was a target of several previous radio monitorings. Long-term  radio observations provided by the Michigan group\footnote{\url{https://dept.astro.lsa.umich.edu/}} during the period of 1970-2010 showed, for instance, that the flux changes at 14.5\,GHz between 0.3-2.8\,Jy, with the highest value archived in 1993. The second prominent outburst reported by the scientists is observed at the end of 2009 with an observed flux level of about 2.4\,Jy. The second mentioned flare can correspond to the main outburst (MJD54900) visible in OVRO observations.

\subsection*{Optical variability}
The optical monitoring with Catalina includes 100\,months of observations, while the monitoring with KAIT includes about 54\,months. The significant optical variability of the source is revealed in several flares that were observed both in R and V band. The strongest outburst took place between MJD55822 and MJD55951, and it shows flux changes in the range of about $3\cdot10^{-11}$\,erg\,cm$^{-2}$\,s$^{-1}$ in V band (during this period \pks\ was not monitored with KAIT). 
Observations of the blazar with KAIT reveal a prominent outburst with 3.5\,mag amplitude. We highlight here that this is the largest optical flare ever reported for this blazar. Previous studies that focused on the optical variability of \pks\ have also shown a significant outburst  of 3\,mag \citep[][]{ Usher74}.

\subsection*{X-ray and UV variability}
\begin{figure}[t]
\centering{\includegraphics[width=0.45\textwidth]{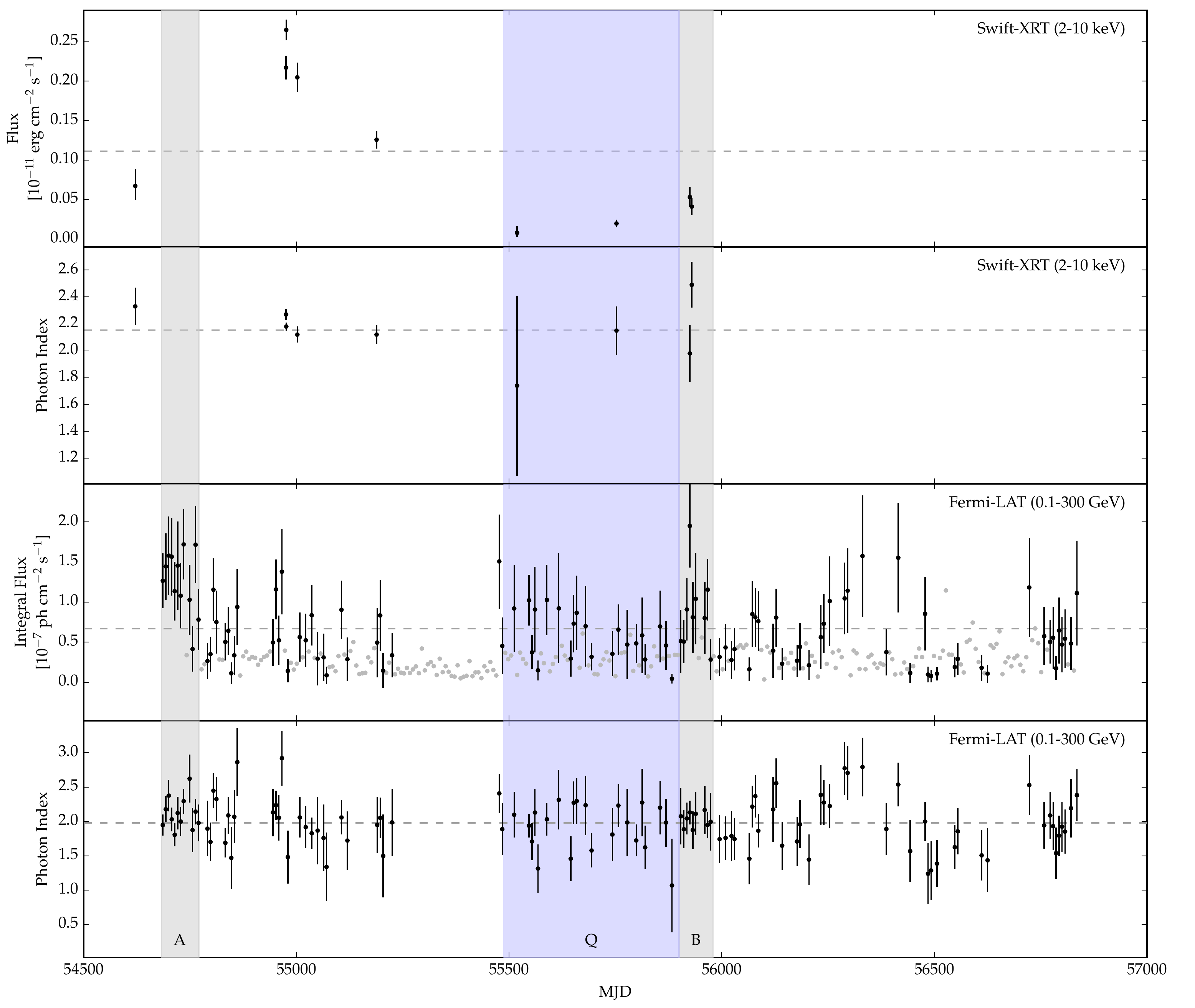}}
\caption[]{Temporal and spectral variability of \pks\ observed with Swift/XRT and  Fermi-/LAT. The following panels present
the X-ray temporal and spectral variability and the $\gamma$-ray temporal and spectral variability.
For Swift/XRT observations one point correspond to one day of observations, while in the case of Fermi-/LAT data one point is for weekly integrated observations. We plot the flux upper
limits n the Fermi/LAT light curve with grey points.
The time periods corresponding to A and B flares are marked in grey, while the quiescence period Q is shown in blue.}
\label{spec_lc}
\end{figure}

Nine pointing observations of \pks\ with \uvot\ and \xrt\ also show changes in the flux. In the X-ray range the largest outburst was observed in June 2009. 
The highest flares observed with \uvot\ occurred in June 2009 and January 2012. The first flare is simultaneous with that of \xrt, while the second optical-UV outburst is not correlated with the X-ray flux. 
No significant variability is observed in the \xrt\ spectral index (see Fig.~\ref{spec_lc}).

\subsection*{GeV temporal and spectral variability}\label{gev_var}
The gamma-ray monitoring of the blazar presented here includes more than five years of observations beginning from the mission start date. The observed flux in this energy range  shows significant temporal variability. The variability is also observed for the spectral index in this wavelength.
We note that in the GeV regime the observed emission is not very strong because the light curve (see Fig.~\ref{all_lc}) includes many of the flux upper limits points and the flux points have large error bars, even though the bin size is set to seven days.

To study variability patterns in the GeV domain in greater detail,
we investigated three intervals that are defined below. 
All the intervals are marked in Fig.~\ref{all_lc}.
\begin{itemize}

\item \textbf{Flare A interval}.
The first flare (hereafter flare A) occurred between $\sim$MJD\,54683 (\textit{Fermi}-LAT mission start date) and MJD\, 54770.1 and is characterized by the elevated flux in GeV, optical, and radio range. There are no observations performed with \xrt\ and \uvot\  during this period, therefore we cannot confirm or to exclude the hypothesis about a high state of the blazar in UV and soft X-rays.

\item \textbf{Flare B interval}.
The second flare (hereafter flare B) is observed between MJD\,55900 and MJD\,55980 in GeV, X-ray, and optical bands. Radio monitoring during this period only shows slight flux oscillation below the mean flux. Surprisingly, two observations with \xrt\ do not show an exceptionally high flux. Short time variations of the elevated flux can be noticed in the UV and optical data obtained with \uvot, Catalina, and KAIT. The shape of the peak in the GeV range shows two separate components that cannot be distinguished in the optical observations in either KAIT or Catalina data.

\item \textbf{The quiescent-state interval}.
The quiescent state of the source has been chosen between MJD\,55450 and MJD\,55830. During this period the flux oscillates around the mean value. The choice of the quiescence state is dictated by two aspects: the flux during the chosen period should not exhibit any significant outbursts, and the length of the interval should be long enough to determine a good-quality spectrum. As mentioned before, large flux uncertainties can suggest variability, but within the error bars the variations are not significant
in the observation period.

\end{itemize}

To study the spectral properties of \pks, we used\ four time periods: two flares A and B, quiescent (low) state (as defined above), and the time period covering all observations.
For each of the defined intervals the photon spectrum was calculated using two spectral models following \cite[e.g.][]{Massaro04}, a single power-law:

\begin{equation}
F(E)=N_p  \left( \frac{E}{E_0}\right)^{-{\Gamma}},
\end{equation}
and a log-parabolic one:
\begin{equation}
F(E)=N_l  \left( \frac{E}{E_0}\right)^{-({\alpha+\beta \log (E/E_0)})}.
\end{equation}
For both models the break energy was set to $E_0=100$\,MeV and
was frozen in the fitting procedure. %\cite[descibe e.g. by][]{Massaro04}.

\begin{figure}[t]
\centering{\includegraphics[width=0.48\textwidth]{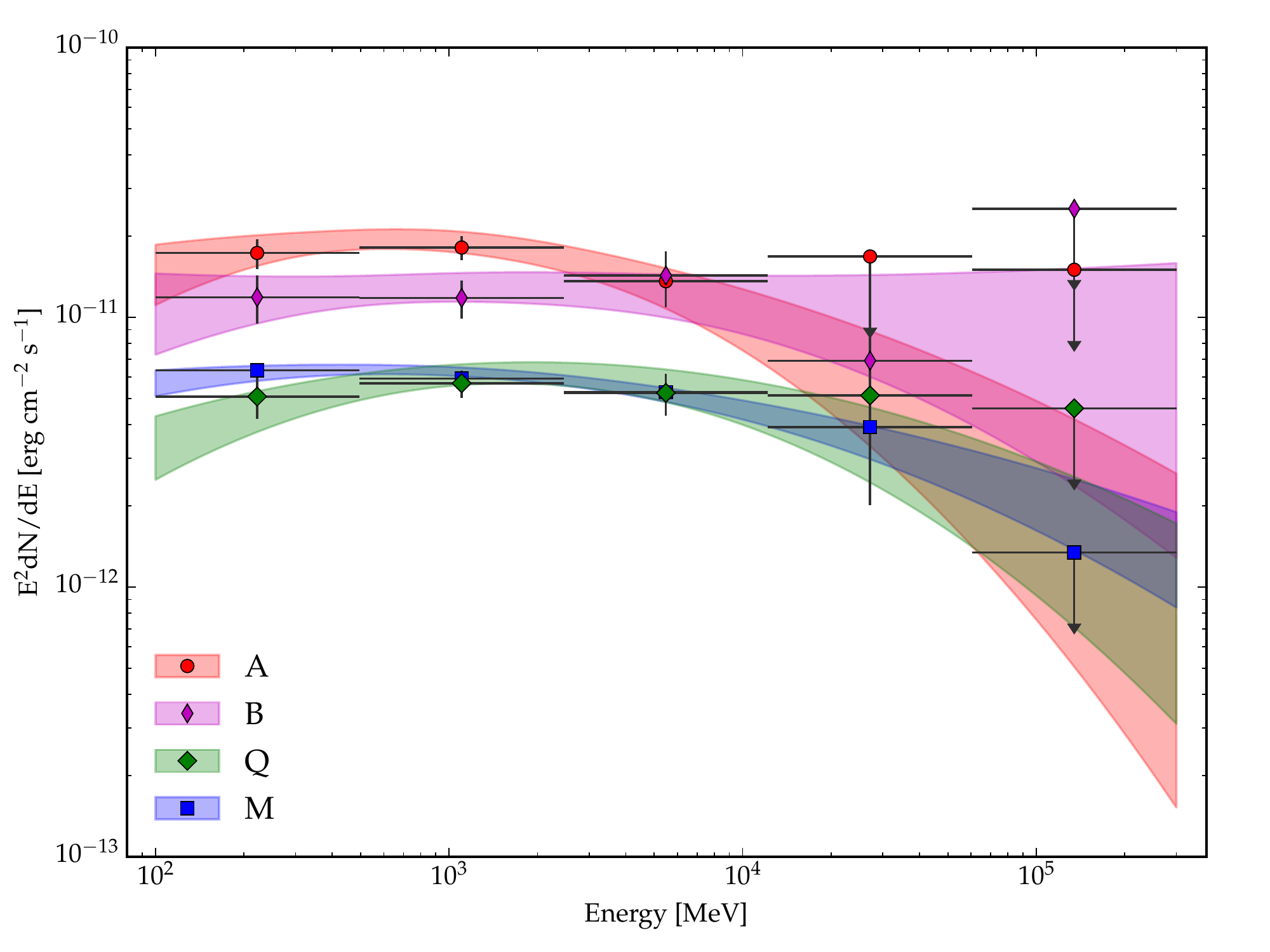}}
\caption[]{Spectral energy distribution for \lat\ data. Data are fitted using log-parabola models. The following colours corresponds to different time epochs: flare A (red), flare B (magenta), quiescent period (Q) defined in Sect.~\ref{gev_var} (green), and all data period (M, blue).  }
\label{sed}
\end{figure}

The spectral points were calculated by dividing the data set into five logarithmically equal energy bins and a separate likelihood analysis was run for each bin. Unfortunately, \pks\ is too faint to obtain flux points in the highest energies in the GeV domain. In this case ($\textrm{TS}<9$), only the flux upper limits were derived. A one-sigma butterfly contour was calculated using the covariance matrix obtained with  the \verb|gtlike| procedure \citep{Abdo09}.
The parameters of the spectral fit and the test statistic (TS) for each model and time interval are collected in Table~\ref{table_lat}. The TS for each interval favours the log-parabolic scenario for the \lat\ data. The log-parabolic fits and the spectral points are shown in the $\nu F_{\nu}$ representation in Fig.~\ref{sed}.
We recall that in the log-parabolic model the $\alpha$ parameter corresponds to the  spectral index, while $\beta$ gives information about the curvature. The spectral parameters collected in Table~\ref{table_lat}  do not show significant spectral variability; $\alpha$ is almost constant within the error bars. The curvature parameter changes, but again uncertainties are too large to confirm or exclude significant changes of this parameter. Furthermore, for the B flare $\beta$ = 0.04 $\pm$ 0.04, which means that here a scenario with zero curvature is also possible, which reduces to a single power-law description.

\subsection{Colour-magnitude relation}
\begin{figure}[t]
\centering{\includegraphics[width=0.48\textwidth]{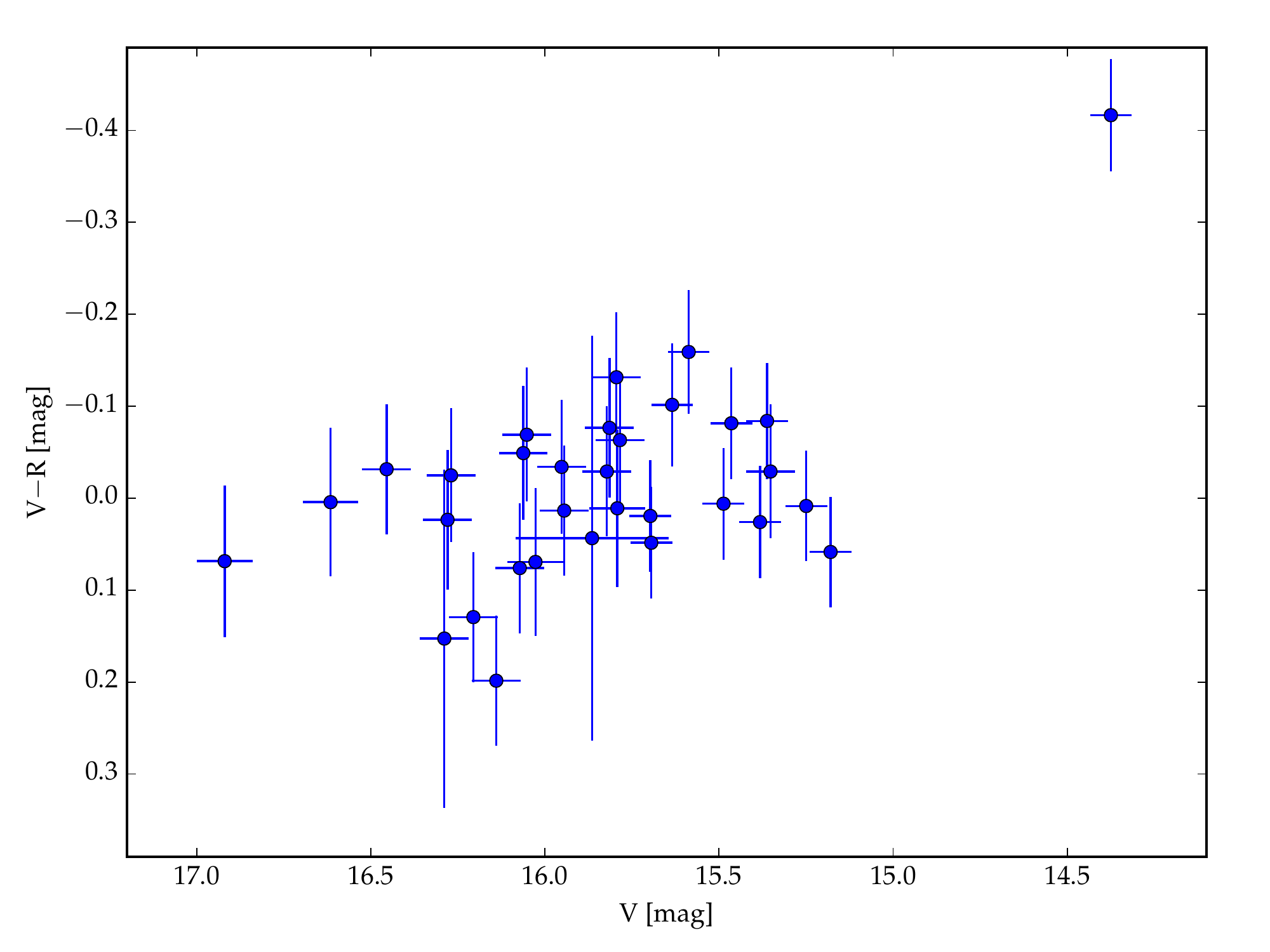}}

\caption[]{Colour-magnitude plot for \pks . Each point corresponds to seven days of binned observations. }
\label{color-mag}
\end{figure}

Figure \ref{color-mag} shows the colour-magnitude diagram for the optical observations of \pks\ obtained with KAIT and Catalina. Neither a bluer-when-brighter nor a redder-when-brighter chromatism is found in data. The result is consistent with those reported by \cite{Ikejiri2011} and \cite{Wierzcholska2014} for this blazar. The Pearson correlation coefficient calculated for the colour-magnitude relation is $0.45\pm0.11$ (see Appendix \ref{appendix:error} for details on how the uncertainty was estimated).
We also note that the optical observations of \pks\ were not corrected for the contribution of the host galaxy or for the contamination from the emission of the accretion disc.

\section{Multi-frequency correlations} \label{sec:correlations}

The simultaneous long-term multi-frequency observations of \pks\ readily allow for correlation studies.
The standard way  to quantify possible relations between two datasets is the discrete correlation function (DCF) following \cite{Edelson88}. Unfortunately, this method  does not work for sparse and non-uniformly sampled light curves.
In this case, the cross-correlation function can be better estimated by the $z$-transformed discrete correlation function (ZDCF). The algorithm has been described in detail by \cite{tal97}.

The ZDCF was calculated for four cases: a comparison of optical data in R and V band, optical data in V band and radio data, optical data in V band and $\gamma$-ray data, and radio data and $\gamma$-ray. To find the peak location for the ZDCF, a maximum-likelihood was calculated for each case using the PLIKE algorithm \citep{tal13}. The peak location,
$\tau_\textrm{max}$, represents the most probable time-lag between the
two light curves.
For each combination of the two light curves, we calculated the following quantities: $\tau_\textrm{max}$, ZDCF$(\tau_\textrm{max})$, and  the Pearson correlation coefficient for the given light curves shifted according to the $\tau_\textrm{max}$. The results are gathered in Table \ref{table_zdcf}.

\begin{figure*}[]
\centering{\includegraphics[width=0.49\textwidth]{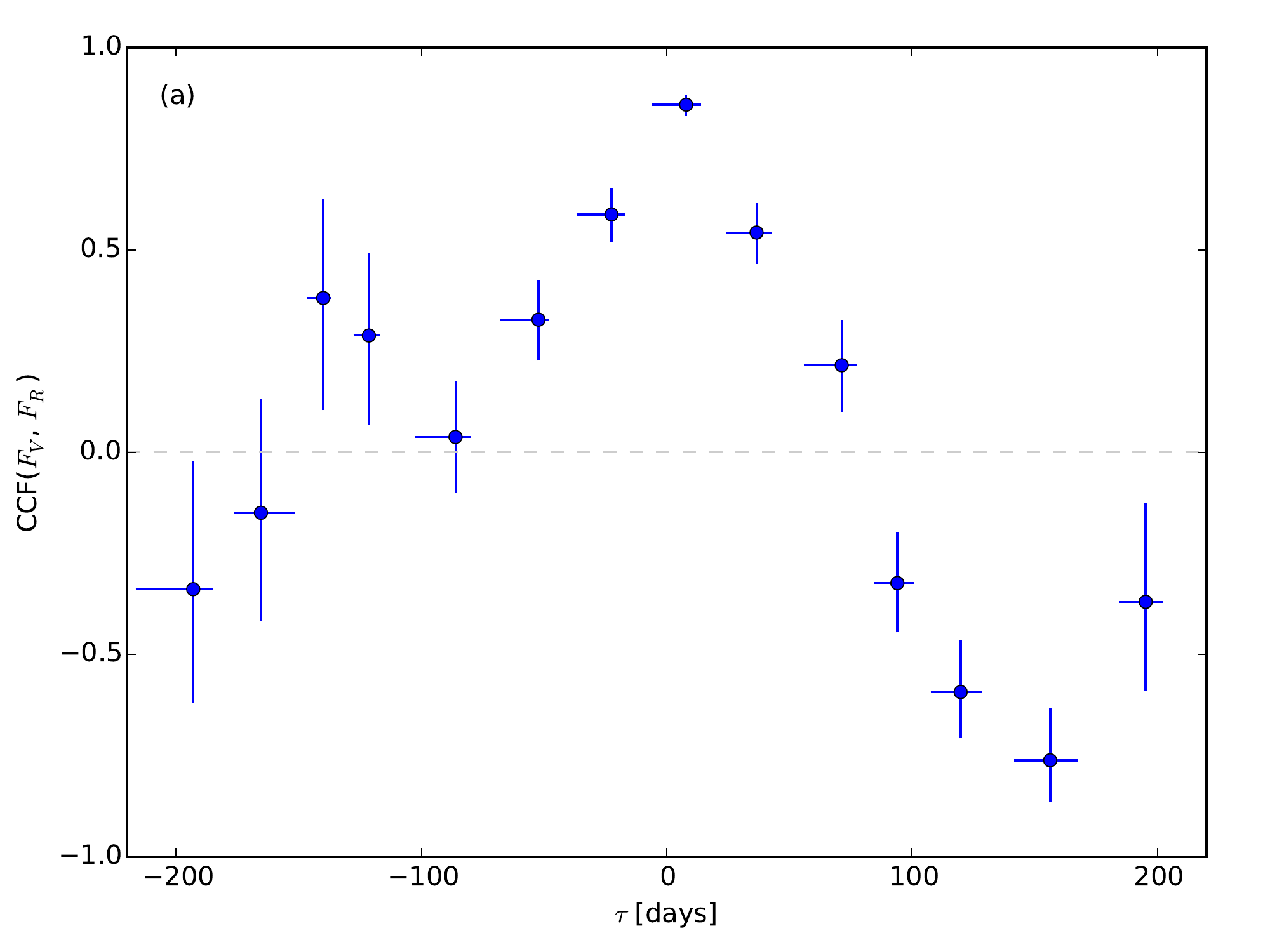}}
\centering{\includegraphics[width=0.49\textwidth]{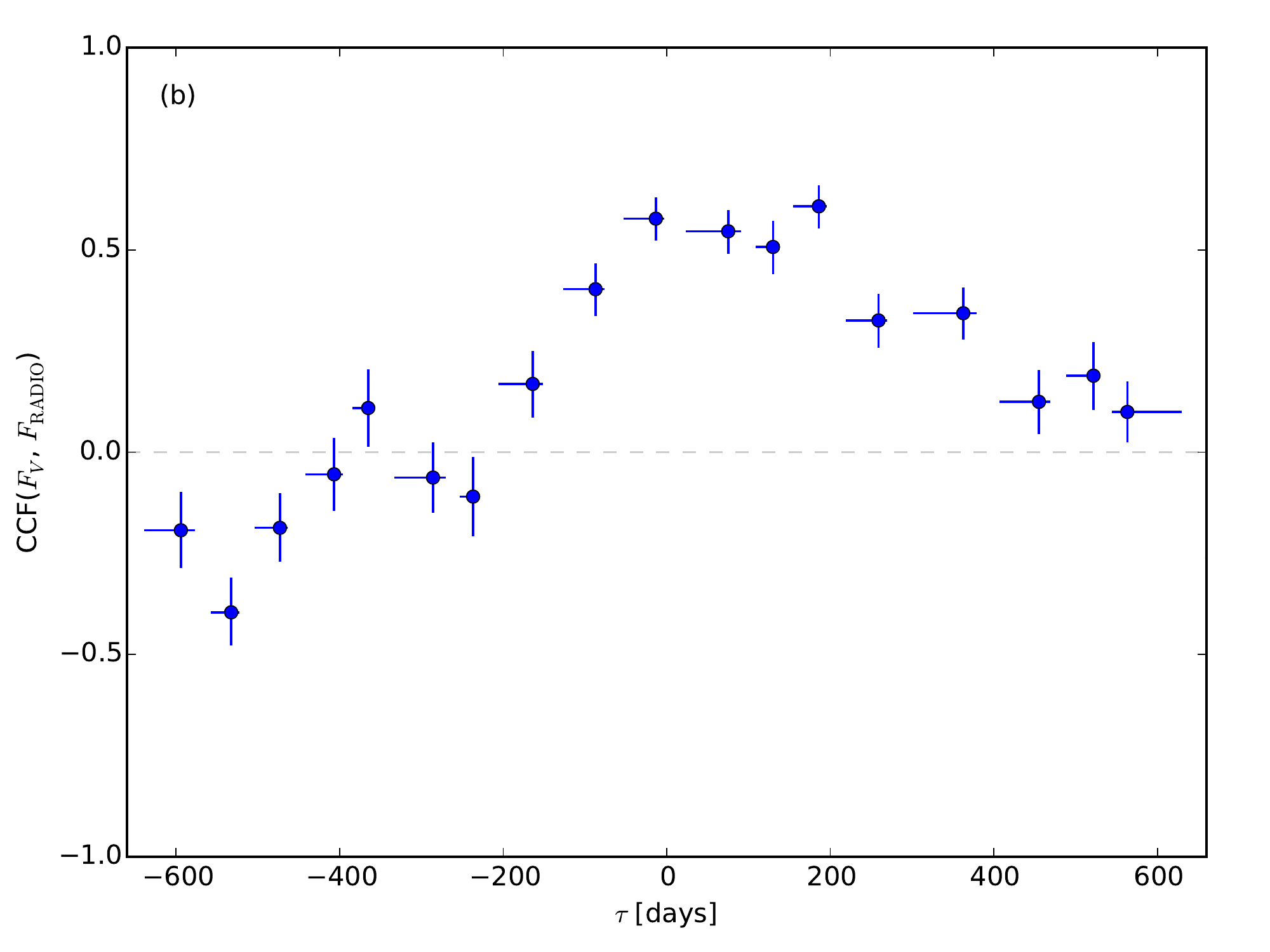}}\\
\centering{\includegraphics[width=0.49\textwidth]{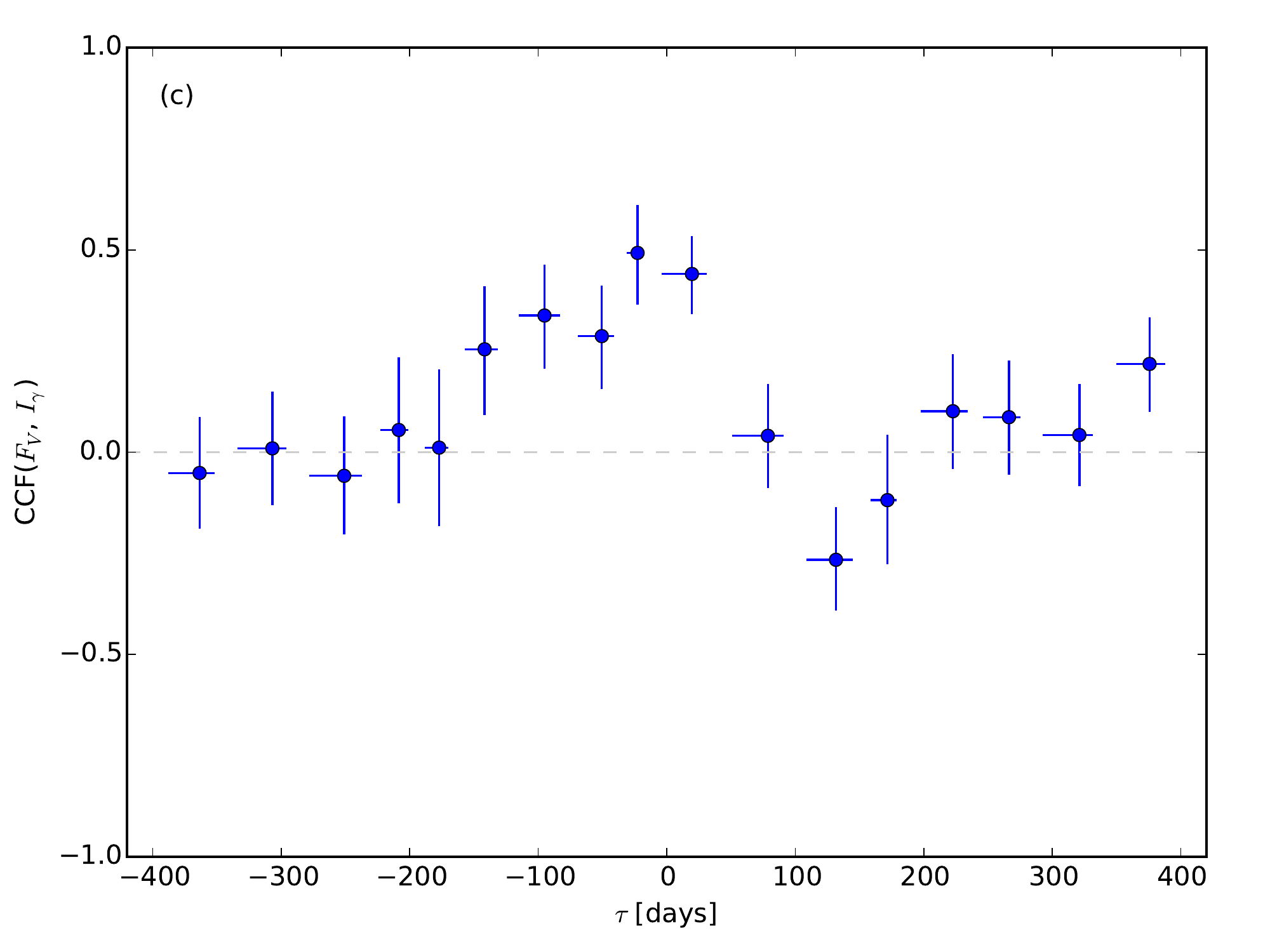}}
\centering{\includegraphics[width=0.49\textwidth]{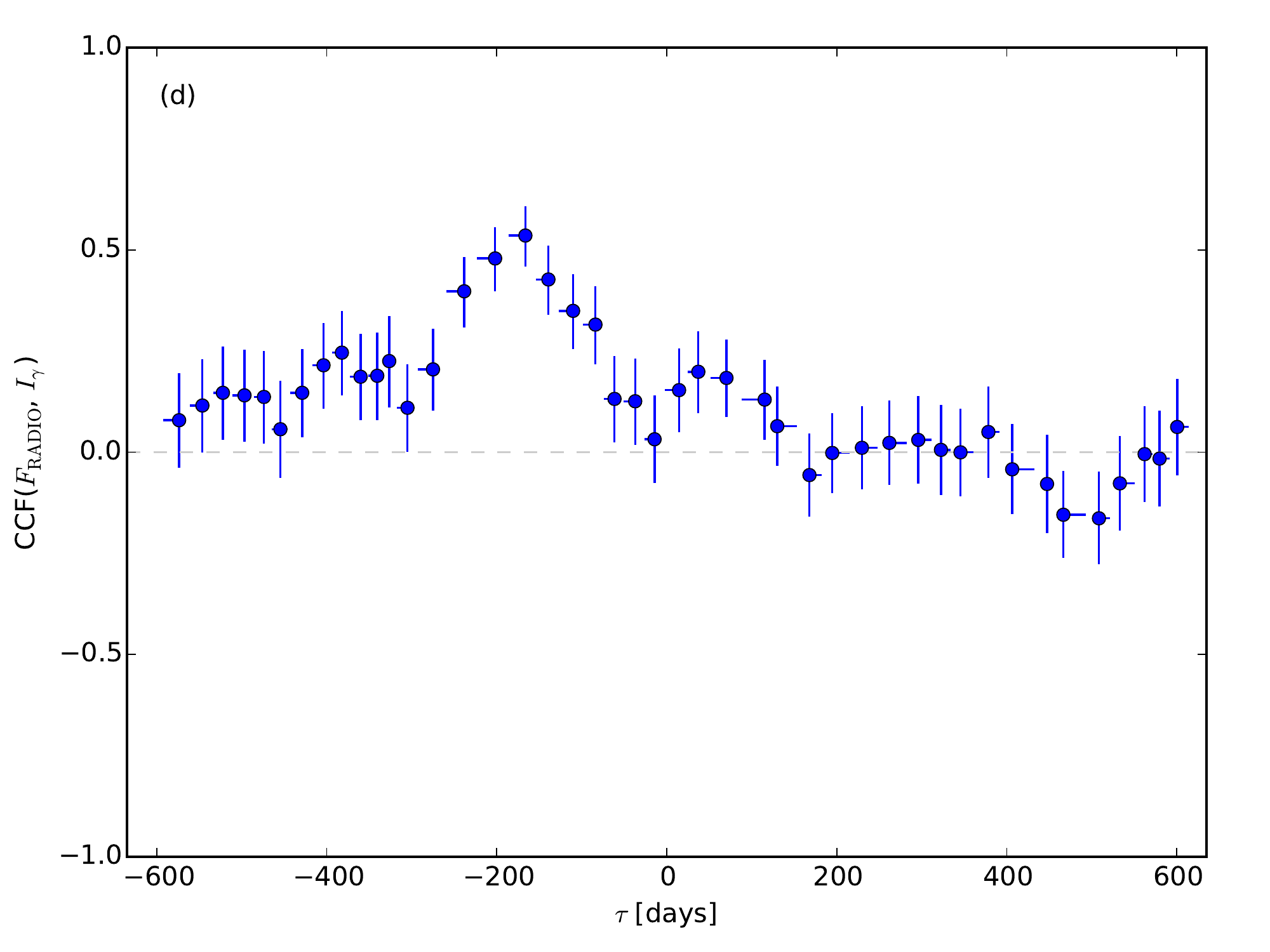}}
\caption[]{Estimate of the cross-correlation function for different time lags for long-term light curve of \pks\ (see Fig.\ref{all_lc}). All data are binned in weekly averaged intervals. The following subplots show the CCF for (a) optical flux in V band, $F_V$ versus optical flux in R band, $F_R$ as a function of time delay; (b) optical flux in V band, $F_V$ versus radio flux $F_\textrm{RADIO}$ as a function of time delay; (c) optical flux in V band, $F_V$ versus $\gamma$-ray integrated flux $I_\gamma$ as a function of time delay; (d) radio flux, $F_\textrm{RADIO}$ versus $\gamma$-ray integrated flux, $I_\gamma$ as a function of time delay. }
\label{ZDCF}
\end{figure*}

In all the cases the flux-flux relation for offset $\Delta t = 0$ and for $\Delta t = \tau_\textrm{max}$ are presented in Fig.~\ref{ZDCF}.  If necessary, the light curves were binned according to the bins of the \lat\ light curves. For the flux-flux relation the Pearson correlation coefficient was calculated, and its error was estimated using a Monte Carlo simulation (see Appendix \ref{appendix:error} for details).
Then the flux-flux relation was fitted with a linear function.

The calculated ZDCF values strongly suggest a correlation between the optical observations in V and R band with a time-lag of zero\,days.
For the other comparisons, no significant correlations for any time-lag were found.  
Radio and optical data show a maximum of the ZDCF function for a time lag of about 200\,days
with a correlation coefficient of  0.5, which does not allow stating it as significant. 
For $\gamma$-ray and optical data the maximum of ZDCF does not exceed 0.5. 
Similar results were also found for the radio and $\gamma$-ray data set.
On the other hand, for the second and fourth case (comparison
of optical - radio and $\gamma$-ray - radio) the Pearson correlation coefficient calculated for the corresponding $\tau_\textrm{max}$ (see Table \ref{table_zdcf}) is about 0.6. This indicates a weak correlation between the emission at the two mentioned wavelengths. We note here that Pearson's correlation coefficient was calculated for binned data in seven-day intervals. This may influence the results because in this case the compared data are only quasi-simultaneous. Moreover, as mentioned before, $\gamma$-ray data have large uncertainties, which weakens the statistical importance of this relation.

\section{Summary and conclusions} \label{summary}
We have presented multi-wavelength monitoring of the blazar \pks\ for nine years, consisting of observations performed in the radio band with OVRO, in the optical and UV bands with KAIT, Catalina and \uvot, in the X-ray band with  \xrt, and in the high energy $\gamma$-ray wavelength with \lat.
It is the longest published monitoring of this source ever. It is also worth mentioning that except in the optical and radio ranges, \pks\ was not studied in detail at the different wavelengths.
During the period studied here, substantial variability is observed in all wavelengths, which in the case of the optical and radio band is consistent with observations reported for example by \cite{Ross70, Carswell73, Wills92, Falomo96, Stickel93, Sefako01}. 
But for the case of other frequencies this is the first time that variability is reported.
We observed 3.5\,mag variations in  the optical band in KAIT data during prominent outbursts, which is the strongest change in the flux for this source ever reported. 
The variability of \pks\ that is observed in the other wavelengths seems to be more complex.

The first flare, called~A here, is characterized by higher flux in the GeV and optical range.
The interpolation of the radio light curve indicates that the observed emission does not show a flare. The second flare discussed in this paper, flare~B, is described by higher flux in the GeV, optical, and UV range. However, neither in radio range nor in X-ray are there any significant changes in the flux level. 
Highly variable emission was quantified using $F_{var}$ values, and it indicates that the highest variability is found in the X-ray range and the lowest in the GeV domain.

The correlation studies of \pks\ show a strong correlation between the optical emission in V and R band and no significant correlations between other wavelengths. The analysis of the cross-correlation function did not show any time lag for which a linear relation could be resolved.
The variability observed in different time scales is a common feature for this class of objects. 
But for the case of the eponymous blazar, the fact that variations are not correlated suggests that the standard, most popular leptonic SSC model might not be enough to describe the emission processes in this object.

The comparison of the optical colour and magnitude of \pks\ does not show either a bluer-when-brighter or a redder-when-brighter relation for the blazar. The lack of such a relation has also been reported by \cite{Ikejiri2011} and \cite{Wierzcholska2014}.
The bluer-when-brighter trend is a common feature of the BL Lac-type blazars, where the optical emission is mostly dominated by the synchrotron radiation from the jet. The lack of such behaviour can be caused by the fact that the correlation is hidden by a few overlapping branches, for which chromatism is present individually. For this case it is possible to distinguish separate states in the colour-magnitude diagram, for which bluer-when-brighter relation is evident. This scenario is possible for \pks\, where the colour-magnitude can disclose substructures with a clear bluer-when-brighter relation. 

The optical data we presented were not corrected for either the contribution of the host galaxy or for the contamination from the accretion disc. \pks\ is a luminous LBL-type blazar (with an optical luminosity of about 10$^{46}$\,erg\,s$^{-1}$) with
an apparently high accretion rate \citep{Sbarrato12}. Not taking into account the accretion disc effects may lead to  redder-when-brighter results in colour-magnitude diagrams, while the neglecting the host galaxy contribution causes a bluer-when-brighter effect. Hence, according to the optical observations of \pks, the lack of any chromatism in the colour-magnitude diagram does not definitely exclude such a relation for the blazar.

The spectral variability studies in GeV energy range show that in this domain the favourable model to describe spectra is the log-parabolic one. 
Previously,  such a model was successfully used in the spectral analysis for other blazars, as reported for instance by
\cite{Massaro04, Massaro04b, Massaro06, Massaro08, Tramacere07, Tramacere09, Tramacere11,  Giommi05, Perri07, Donato05}.

The studies on the long-term emission in \pks\ confirm the importance of the simultaneous multi-frequency monitoring of blazars in
both the flaring and quiescence states. Our results strongly encourage further monitoring of \pks\ in the multi-frequency simultaneous campaigns.

%--------------------------
%--------------------------

 \begin{table*}  
\caption[]{Parameters of the spectral analysis of \textit{Swift}/XRT data. } 
\centering
\begin{tabular}{c|c|c|c|c|c}
\hline
\hline
Observation ID & Observation date & Exposure  &  $F_{2-10\,\textrm{keV}}$  &   $\Gamma$ & $\chi^2_{red}/n_{d.o.f.}$    \\
                &                    &(ks)&(10$^{-12}$\,erg\,cm$^{-2}$\,s$^{-1}$) &  &   \\
      (1)          & (2) & (3) & (4) & (5) & (6) \\
 
 \hline
 
 36364001         & 04/06/2008 & 0.8  & $1.60\pm0.18$ & $2.33\pm0.14$ & 0.645/10   \\
 38093001         & 24/05/2009 & 5.7  & $4.13\pm0.04$ & $2.27\pm0.04$ & 1.029/121  \\
 36364002         & 25/05/2009 & 9.7  & $5.42\pm0.03$ & $2.18\pm0.03$ & 0.905/179  \\
 36364003         & 20/06/2009 & 2.6  & $4.39\pm0.06$ & $2.12\pm0.06$ & 0.879/61   \\
 36364004         & 23/12/2009 & 5.0  & $2.23\pm0.08$ & $2.12\pm0.07$ & 1.045/54   \\
 41714001         & 18/11/2010 & 1.6  & $0.54\pm0.03$ & $1.74\pm0.67$ & 0.542/4    \\
 38093002         & 10/07/2011 & 3.8  & $0.32\pm0.03$ & $2.15\pm0.18$ & 0.415/6    \\
 38093003         & 30/12/2011 & 1.5  & $1.05\pm0.26$ & $1.98\pm0.21$ & 1.40/7     \\
 38093004         & 03/01/2012 & 1.4  & $0.86\pm0.19$ & $2.49\pm0.17$ & 1.303/11   \\
 All observations &            & 32.1 & $3.20\pm0.02$ & $2.22\pm0.02$ & 1.091/252  \\
  
\hline
\hline
     
\end{tabular}
\tablefoot{The following columns report (1) the observation ID, (2) the observation time, (3) the exposure of analysed observations,  (4) the integrated flux in the energy range from 2 to 10\,keV, (5) the photon index for the power law fit to the spectrum, (6) the reduced $\chi^2$ and the number of degrees of freedom for the power-law fit. The given $F_{2-10\,\textrm{keV}}$ fluxes are not corrected for Galactic absorption, since in energy band of 2-10\,keV this effect is found to be negligible.}
\label{table_xrt}
\end{table*}

\begin{table*}  
\caption[]{Magnitudes for different epochs from \textit{Swift}/UVOT data for V, B, U, UVW1, UVM2, and UVW2 filters.}
\centering
\begin{tabular}{c|c|c|c|c|c |c}
\hline
\hline
 Observation ID & V & B &  U &  UVW1 & UVM2 & UVW2     \\
 
 \hline
36364001 & --             & --             & --             & --             & --             & $15.13\pm0.06$ \\

38093001 & $15.49\pm0.07$ & $15.91\pm0.06$ & $15.12\pm0.06$ & $15.07\pm0.06$ & $15.14\pm0.07$ & $15.25\pm0.06$ \\

36364002 & --             & --             & $15.09\pm0.05$ & --             & --             & --             \\

36364003 & --             & --             & --             & --             & $15.64\pm0.07$ & --             \\

36364004 & $16.39\pm0.10$ & $17.00\pm0.09$ & $16.03\pm0.08$ & $16.05\pm0.08$ & $16.04\pm0.09$ & $16.23\pm0.07$ \\

41714001 & $16.23\pm0.10$ & $16.67\pm0.09$ & $15.84\pm0.09$ & $15.98\pm0.09$ & $16.02\pm0.10$ & $16.21\pm0.08$ \\

38093002 & $15.67\pm0.07$ & $16.10\pm0.06$ & $15.32\pm0.05$ & $15.33\pm0.07$ & $15.33\pm0.07$ & $15.49\pm0.06$ \\

38093003 & $15.14\pm0.06$ & $15.54\pm0.05$ & $14.75\pm0.06$ & $14.81\pm0.06$ & $14.76\pm0.07$ & $15.04\pm0.06$ \\

38093004 & $14.69\pm0.05$ & $15.05\pm0.05$ & $14.21\pm0.05$ & $14.22\pm0.06$ & $14.24\pm0.06$ & $14.39\pm0.06$ \\

\hline
\hline

\end{tabular}
\tablefoot{The magnitudes are corrected for Galactic extinction.\\ (--) No observation taken in this filter for the given observation ID.}
\label{table_uvot}
\end{table*}

%--------------------------
%--------------------------

 \begin{table*}  
\caption[]{Fractional variability in different energy bands. } 
\centering
\begin{tabular}{c|c|c|c|c}
\hline
\hline
Instrument & Energy band/filter & $F_{var}$ &$\chi^2$/$n_{d.o.f.}$  & Bin size\\
(1) & (2) & (3) & (4) & (5)  \\

\hline
\lat\    & 0.1 -- 300\,GeV & $0.340 \pm 0.073$ & 532/109    & 7\,days\\

\xrt\    & 2 -- 10\,keV    & $0.850 \pm 0.004$ & 876/7      & --\\

\uvot    & UVW2            & $0.597 \pm 0.020$ & 1225/5     & -- \\

\uvot    & UVM2            & $0.654 \pm 0.017$ & 1449/5     & --\\

\uvot    & UVW1            & $0.651 \pm 0.029$ & 665/4      & -- \\

\uvot    & U               & $0.590 \pm 0.019$ & 1022/5     & --\\

\uvot    & B               & $0.640 \pm 0.017$ & 1394/4     & --\\

\uvot    & V               & $0.590 \pm 0.022$ & 642/4      & --\\

Catalina & V               & $0.500 \pm 0.039$ & 29830/316  & 1\,day \\

KAIT     & R               & $0.455 \pm 0.003$ & 349399/164 & 1\,day\\

OVRO     & 15\,GHz         & $0.495 \pm 0.002$ & 196843/377 & 1\,day \\

\hline
\hline
\end{tabular}
\tablefoot{The following columns present (1) the name of the instrument, (2) the energy band or filter, (3) the fractional variability, (4) the chi square value and the number of degrees of freedom for the fit with a constant, (5) the size of the data bins. In the case of \xrt\ and \uvot\ data due to small number of pointing observations the bin sizes are not provided.}
\label{table_fvar}
\end{table*}

\begin{table*}  
\caption[]{Summary of $z$-transformed discrete correlation function.}
\centering
\begin{tabular}{c|c|c|c|c|c}
\hline
\hline
energy bands             & $\tau_\textrm{max}$ & time interval & ZDCF$(\tau_\textrm{max})$ & probability & $R_{bin}$ \\
                         & [days]              &               &                 &             & \\
 (1)                     & (2)                 & (3)           & (4)             & (5)         & (6) \\
 \hline
 V -- R                  & 7.9                 & (-6.0;+20.4)  & $0.860\pm0.026$ & 77$\%$      & $0.95\pm0.11$ \\
 V -- radio              & 186.0               & (+7; +217)    & $0.608\pm0.055$ & 52$\%$      & $0.57\pm0.06$ \\
 V -- $\gamma$-ray       & $-$23.0             & (-88; +21)    & $0.493\pm0.012$ & 44$\%$      & $0.47\pm0.15$ \\
 radio -- $\gamma$-ray   & $-$166.0            & (-200;+148)   & $0.536\pm0.074$ & 46$\%$      & $0.59\pm0.08$ \\

\hline
\hline
     
\end{tabular}
\tablefoot{The following columns present (1) the energy bands for which ZDCF is calculated, (2) the calculated time lag, (3) the time interval for which the time lag is calculated, (4) the ZDCF value for $\tau_\textrm{max}$, (5) the probability for ZDCF, (6) the Pearson correlation coefficient.}
\label{table_zdcf}
\end{table*} 

%--------------------------
%--------------------------

%--------------------------
%--------------------------

\begin{figure*}[t]
\centering{\includegraphics[width=0.49\textwidth]{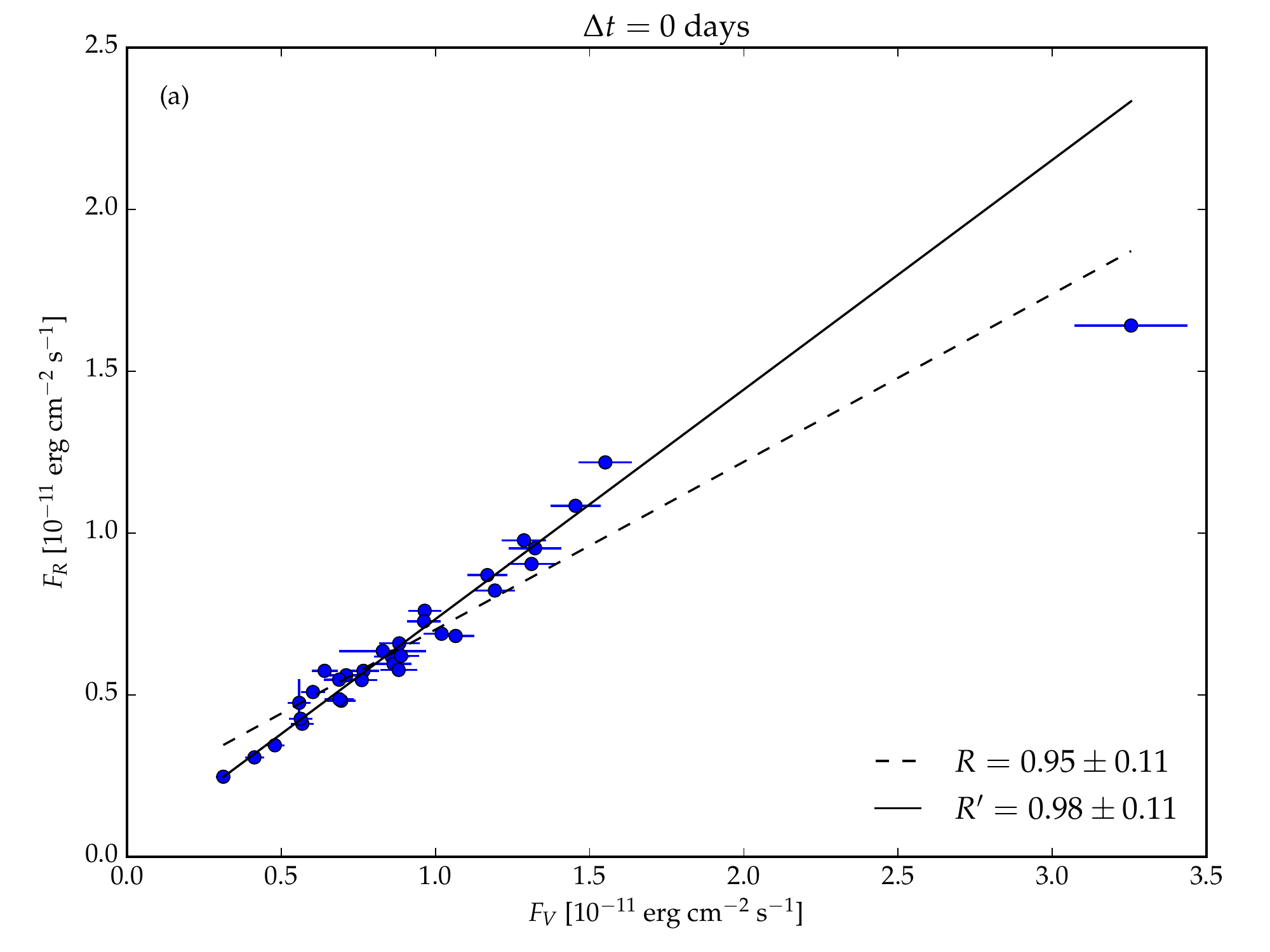}}
\centering{\includegraphics[width=0.49\textwidth]{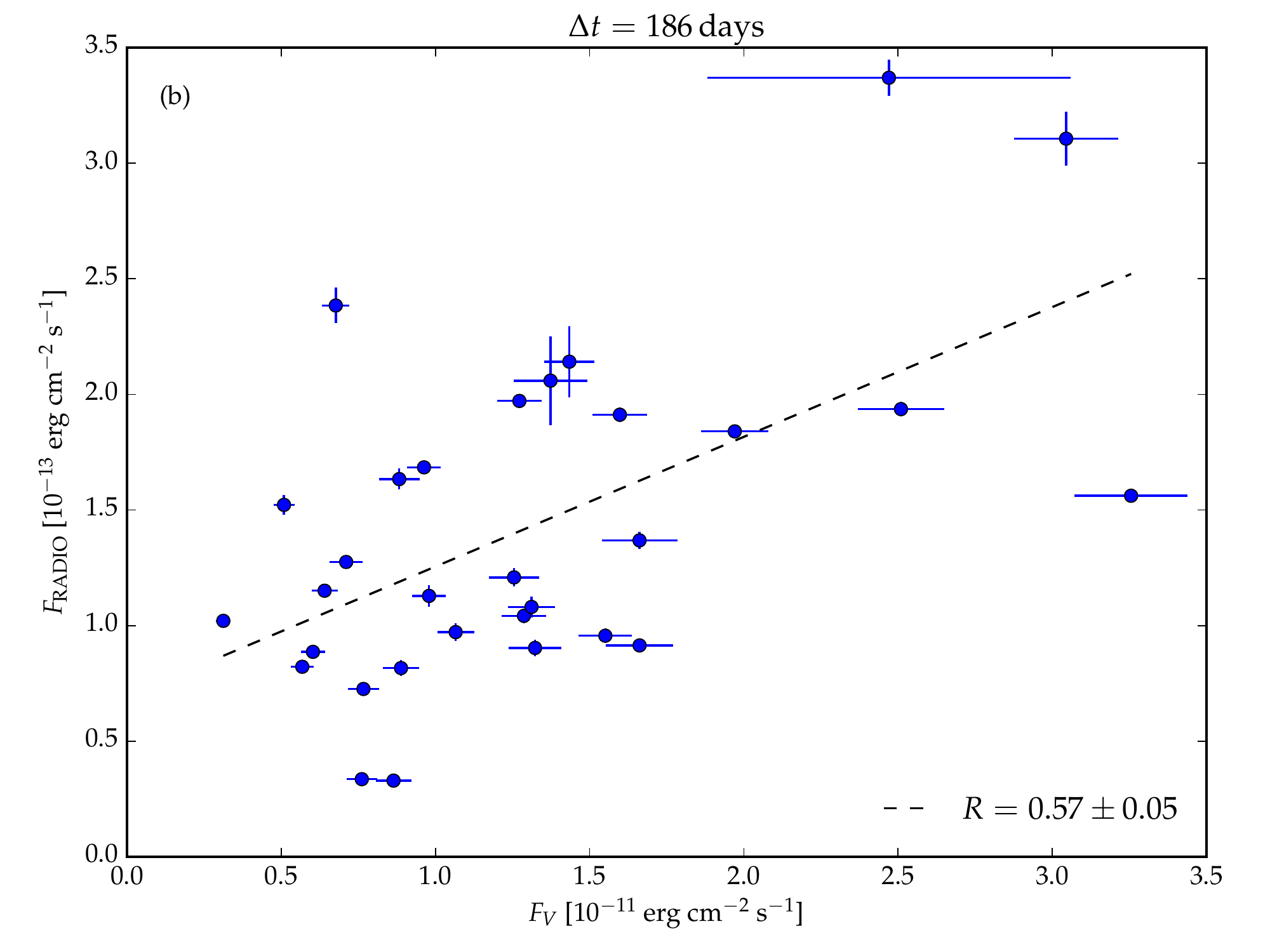}} \\
\centering{\includegraphics[width=0.49\textwidth]{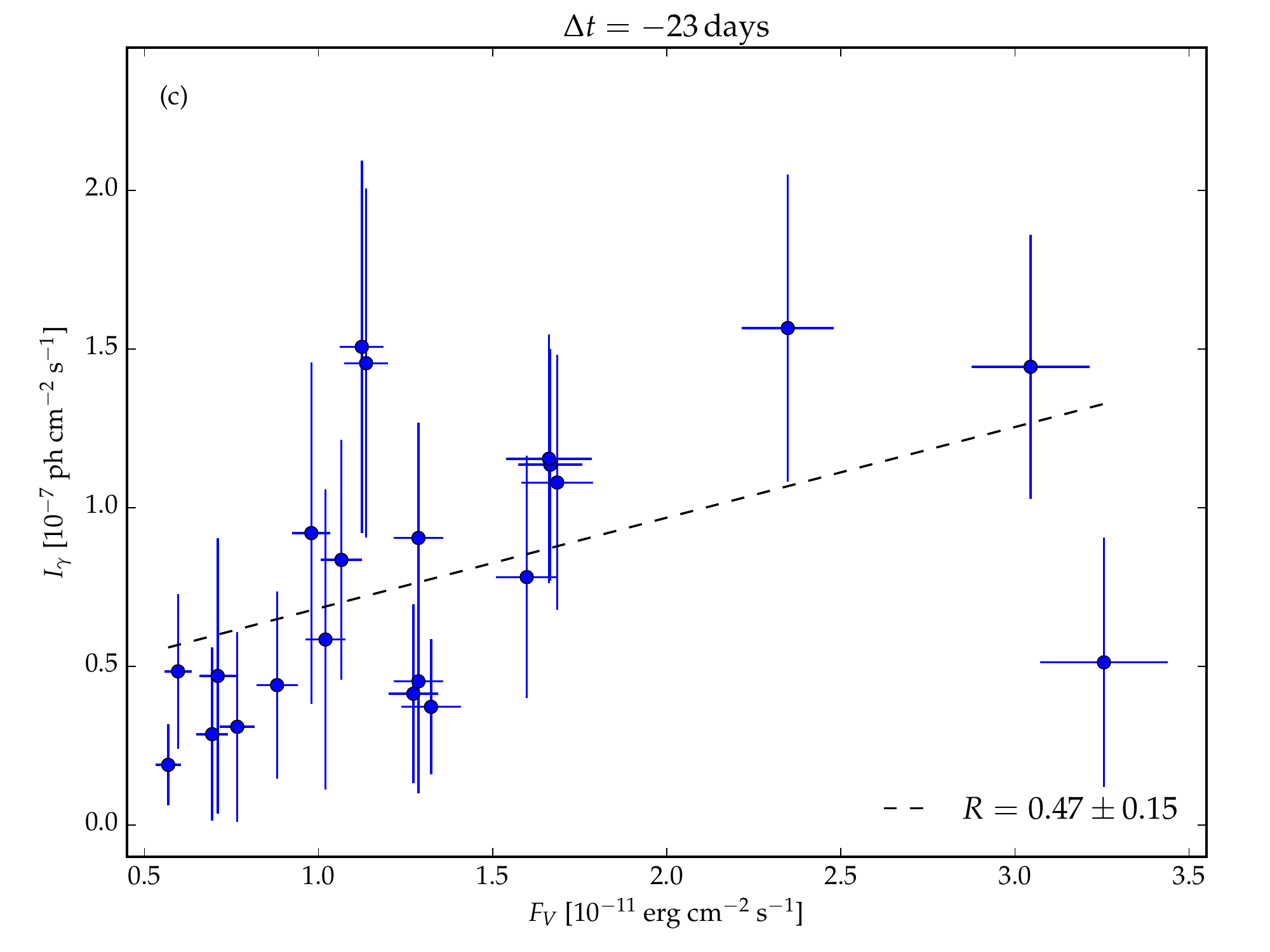}}
\centering{\includegraphics[width=0.49\textwidth]{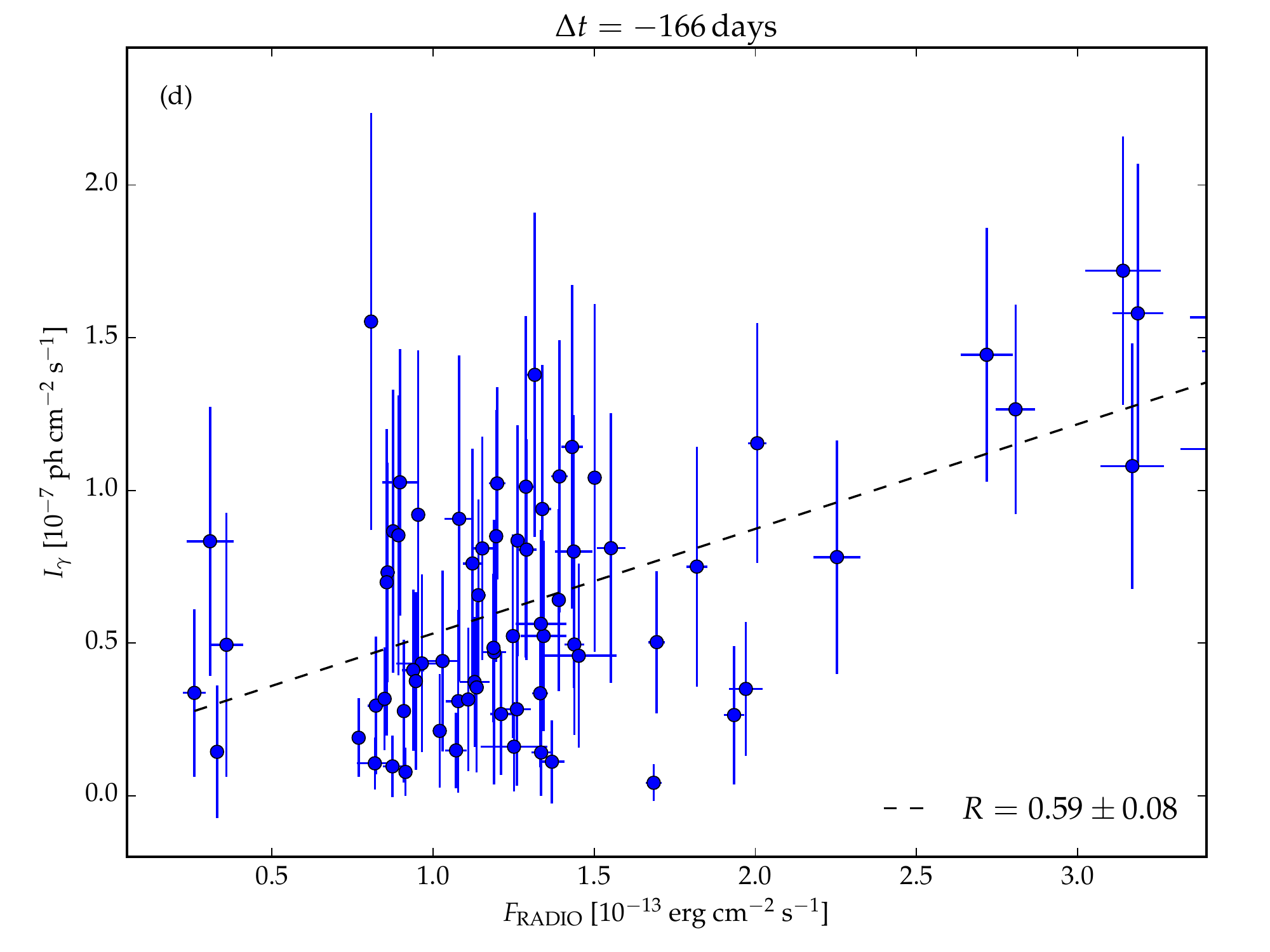}}

\caption[]{Subplot (a) shows the comparison of the optical flux in R ($F_R$) and V ($F_V$) band for \pks\ for simultaneous observations. The Pearson correlation coefficients for the two cases of a full set of data - solid line and without one data point (in right upper corner in the plot) - dashed line, are provided in the lower right corner. The following subplots show the relation for shifted light curves for (b) the radio flux, $F_\textrm{RADIO}$ vs $F_V$, (c) integrated gamma-ray flux, $I_\gamma$ vs $F_V$,  and (d) $I_\gamma$ vs $F_\textrm{RADIO}$. The light curves are shifted by $\Delta t$ according to $\tau_\textrm{max}$ shown in Table. \ref{table_zdcf}. In the lower right corner the Pearson correlation coefficients are provided. }
\label{correlation_all}
\end{figure*}
%--------------------------

\begin{table*}  
\caption[]{Power-law and log-parabolic fit parameters to \lat\ data.}
\centering
\begin{tabular}{c|c|c|c|c|c}
\hline
\hline
 & TS & N (10$^{-1}$ cm$^{-2}$ s$^{-1}$ MeV$^{-1}$) & $\gamma$ & $\alpha$ & $\beta$ \\
 & (1) & (2) & (3) & (4) & (5) \\
  \hline
All data -- power law         & 2290 & $ 4.54 \pm 0.33$ & $2.09 \pm 0.03$ & --              & -- \\
All data -- log-parabola      & 2293 & $ 3.55 \pm 0.39 $  & --              & $1.86 \pm 0.07$ & $0.04 \pm 0.01$ \\

Flare A -- power law          & 555  & $ 13.5\pm 1.9 $ & $2.11 \pm 0.06$ & --              & -- \\
Flare A -- log-parabola       & 559  & $ 9.0\pm 2.3 $  & --              & $1.67 \pm 0.22$ & $0.09 \pm 0.04$ \\

Flare B -- power law          &  304 & $ 7.9\pm 1.8 $ & $2.02 \pm 0.08$ & --              & -- \\
  Flare B -- log-parabola     & 304  & $ 6.4\pm 2.2 $  & --              & $1.82 \pm 0.25$ & $0.04 \pm 0.04$ \\

Quiescent state -- power law  & 379  & $ 3.51 \pm 0.70  $ & $2.01 \pm 0.07$ & --              & -- \\
Quiescent state -- log-parabola &  382 & $ 2.04 \pm 0.56 $  & --              & $1.55 \pm 0.17$ & $0.08 \pm 0.03$ \\

\hline
\hline
     
\end{tabular}
\tablefoot{The following columns present (1) the test statistic for a given fit, (2) the fit normalization: $N = N_p$ for power
law or $N = N_l$ for log-parabola, (3) the spectral index for the power-law fit, (4) the slope  parameter for the log-parabolic fit, (5) the curvature parameter forthe  log-parabolic fit. For all power-law and log-parabolic models the break energy  was a frozen parameter with 100\,MeV.}
\label{table_lat}
\end{table*} 

%--------------------------
%--------------------------

\section*{Acknowledgements}
The author thanks the anonymous referee for constructive
comments that greatly improved the manuscript.
A.W. acknowledges support by Polish Ministry of Science and Higher Education in Mobility Plus Program.

This research has made use of data from the OVRO 40-m monitoring program
\citep{Richards11}, which is supported in part by NASA grants NNX08AW31G and NNX11A043G, and NSF grants AST-0808050 and AST-1109911.
The CSS survey is funded by the National Aeronautics and Space
Administration under Grant No. NNG05GF22G issued through the Science
Mission Directorate Near-Earth Objects Observations Program.  The CRTS
survey is supported by the U.S.~National Science Foundation under
grants AST-0909182 and AST-1313422.
The author gratefully acknowledges the optical, X-ray, and $\gamma$-ray observations provided by the KAIT, \textit{Swift} and \textit{Fermi} teams, respectively.

This research was supported in part by PLGrid Infrastructure. The plots presented in this paper are rendered using Matplotlib \citep{matplotlib}.

\begin{appendix}
\section{Error estimation for the Pearson correlation coefficient}
 \label{appendix:error}
To estimate the uncertainty of the Pearson correlation coefficient, one can use
the Monte Carlo approach.  
Here we assumed that  a
set of points $A = \{(x_i,y_i)\}$ is given and each of the point has its own
corresponding uncertainty values $(\Delta x_i,\Delta y_i)$. In the first step,
for each point new coordinates were drawn randomly according to the normal
distribution for which the mean was set to $x_i$ (or $y_i$) and the standard
deviation to $\Delta x_i$ (or $\Delta y_i$).  This results in a new set of points
$A'$ , and its Pearson correlations coefficient is $C'$.

\begin{figure}
\includegraphics[width=0.45\textwidth]{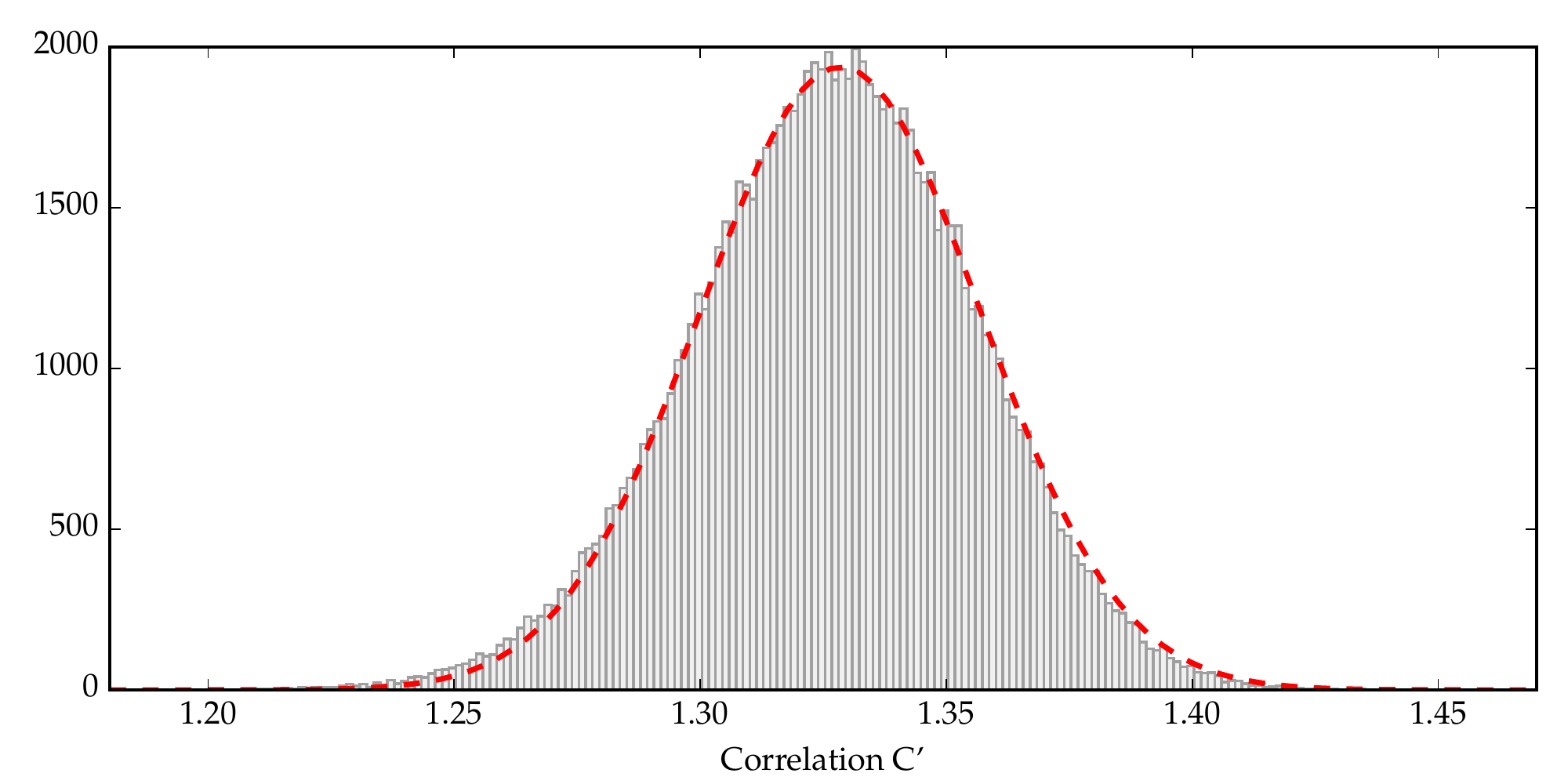}
\caption{Histogram of the Monte Carlo-generated set of $10^5$ Pearson correlation coefficients $\{C'\}$. The red
line shows the Gaussian fit. The $C'$ values were transformed with
the Fisher transformation. The reverse Fisher transformation of the fit gives
a Pearson correlation coefficient equal to 0.87, and its error is estimated
to 0.03.}
\label{fig:gauss}
\end{figure}

Repeating the procedure $N$ times gives a set of Pearson coefficients
$\{C'_n\}$.  If $N$ is large enough, a histogram of the $\{C_n'\}$ should have
roughly a Gaussian shape.  However, because the Pearson coefficient only has values in the range $[-1,1]$ it is good to apply a~Fisher transformation (see
below) on each of the $C'$ value before making the histogram. An example of
such a histogram is shown in Figure~\ref{fig:gauss}.

To find the uncertainty of the Pearson coefficient, a Gaussian function was
fitted to the histogram. The standard deviation of this fit can be used as an
estimate of the Pearson coefficient uncertainty of the original set of
points $A$.  The value found by fitting should be transformed back by the
reverse Fisher transformation.

\subsection*{Fisher transformation}
The Fisher transformation allows representing values that span from $[-1,1]$
in a range of $(-\infty,\infty)$. It is defined as follows:

\begin{equation}
z = \frac{1}{2} \ln \left( \frac{1+r}{1-r} \right) = \atanh r,
\end{equation}
and the reverse:

\begin{equation}
    r = \frac{\exp (2z) - 1}{\exp (2z) + 1}.
\end{equation}

\end{appendix}
 
\bibliographystyle{aa}
\bibliography{bib}

\clearpage

\end{document}